\documentclass[prl,amsmath,preprintnumbers,amssymb,aps,superscriptaddress,twocolumn]{revtex4-1}
\usepackage[utf8]{inputenc}
\usepackage[T1]{fontenc} 
\usepackage{amsmath,amsfonts,amssymb,amsthm,graphics,graphicx,epsfig,times,bbm}
\usepackage[colorlinks=true,citecolor=blue,linkcolor=red]{hyperref}
\usepackage[usenames]{color}
\usepackage{subfigure}
\usepackage{dcolumn}
\usepackage{bm}
\usepackage{color}
\usepackage[dvipsnames]{xcolor}
\usepackage{verbatim}
\usepackage{epstopdf}
\usepackage{amstext}
\usepackage{latexsym}
\usepackage{hyperref}
\usepackage{psfrag}
\usepackage{xcolor}
\usepackage{titlesec}
\usepackage{fullpage}
\usepackage{float}
\usepackage{tikz}
\usepackage{ stmaryrd }
\usetikzlibrary{calc}
\usepackage{tabularx}
\usepackage[colorinlistoftodos]{todonotes} 

\def\l{{\it  l}}

\def\beq{\begin{equation}}
\def\eeq{\end{equation}} 

\newcommand{\beqa}{\begin{eqnarray}}
\newcommand{\eeqa}{\end{eqnarray}}

\newcommand{\bra}[1]{\langle #1\rvert}
\newcommand{\ket}[1]{\lvert#1\rangle}

\newcommand{\Sz}[2][]{{} #2^{\dagger #1}}

\def\Sz{\hat{\sigma}_z}
\def\Sy{\hat{\sigma}_y}
\def\Sx{\hat{\sigma}_x}
\def\Sp{\hat{\sigma}_+}
\def\Sm{\hat{\sigma}_-}
\def\Tz{\hat{\tau}_z}
\def\Ty{\hat{\tau}_y}

\def\imm{\mathrm{i}}

\def\xop{\hat{x}}
\def\pop{\hat{p}}
\def\aop{\hat{a}}
\def\adag{{\hat{a}}^{\dagger}}
\def\rop{\hat{\rho}}

\newcommand{\Round}[1]{\left( #1\right)}
\newcommand{\Square}[1]{\left[ #1\right]}
\newcommand{\Curly}[1]{\left\{ #1\right\}}

\makeatletter
\renewcommand{\boxed}[2]{\textcolor{#1}{%
\tikz[baseline={([yshift=-1ex]current bounding box.center)}] \node [rectangle, minimum width=1ex,rounded corners,draw] {\normalcolor\m@th$\displaystyle#2$};}}
 \makeatother

\titleformat{\section}[hang]{\filcenter\normalfont \bfseries}{S\arabic{section} }{0.1cm}{\bfseries}
\titleformat{\subsection}[hang]{\normalfont \bfseries}{S\arabic{section}.\letter{subsection} }{0.1cm}{\bfseries}

\begin{document}

\title{Critical Quantum metrology with a finite-component quantum phase transition}
\date{\today}
\author{Louis Garbe}
\affiliation{Laboratoire Mat\'eriaux et Ph\'enom\`enes Quantiques, Universit\'e Paris Diderot, CNRS UMR 7162, Sorbonne Paris Cit\`e, France}
\author{Matteo Bina}
\affiliation{Quantum Technology Lab, Dipartimento di Fisica {\em Aldo Pontremoli}, Universit\`a degli Studi di Milano, I-20133 Milano, Italy}
\author{Arne Keller}
\affiliation{Laboratoire Mat\'eriaux et Ph\'enom\`enes Quantiques, Universit\'e Paris Diderot, CNRS UMR 7162, Sorbonne Paris Cit\`e, France}
\affiliation{Universit\'e Paris-Sud, Universit\'e Paris-Saclay, France}
\author{Matteo G. A. Paris}
\affiliation{Quantum Technology Lab, Dipartimento di Fisica {\em Aldo Pontremoli}, Universit\`a degli Studi di Milano, I-20133 Milano, Italy}
\author{Simone Felicetti}
\affiliation{Departamento de F\'isica Te\`orica de la Materia Condensada 
and Condensed Matter Physics Center (IFIMAC), Universidad Aut\`onoma de 
Madrid, E-28049 Madrid, Spain}
\begin{abstract}
Physical systems close to a quantum phase transition exhibit 
a divergent susceptibility, suggesting that an arbitrarily-high
precision may be achieved by exploiting quantum critical systems  
as  probes to estimate a physical parameter. However, such an improvement in sensitivity is counterbalanced by the closing of the energy gap, which implies a critical slowing down
and an inevitable growth of the protocol duration. Here, we design different metrological protocols that make use of the superradiant phase transition of the quantum Rabi model, a finite-component system composed of a single two-level atom interacting with a single bosonic mode. We show that, in spite of the critical slowing down, critical quantum optical systems can lead to a quantum-enhanced time-scaling of the quantum Fisher information, and so of the measurement sensitivity. 
\end{abstract}
\maketitle

In a system close to a critical point, small variations of physical parameters may lead to dramatic changes in the equilibrium state properties. The possibility of exploiting this sensitivity for metrological purposes is well known, and it has already been applied in classical devices, e.g. in superconducting  transition-edge sensor \cite{irwin_transition-edge_2005}.  Besides, the development of quantum metrology has extensively shown that quantum  states can outperform their classical counterparts for sensing tasks \cite{demkowicz-dobrzanski_chapter_2015}. Therefore, a question naturally arises: what sensitivity can be achieved using interacting systems close to a quantum-critical point? In the last few years, this question has attracted growing interest and it has been addressed by different methods \cite{zanardi_quantum_2008,ivanov_adiabatic_2013,tsang_quantum_2013,bina_dicke_2016,macieszczak_dynamical_2016,fernandez-lorenzo_quantum_2017,rams_at_2018}. These studies may be roughly divided in two classes. 

The first approach, which we will call the "dynamical" paradigm \cite{tsang_quantum_2013,macieszczak_dynamical_2016}, focus on the time evolution induced by a Hamiltonian close to a critical point. In this approach, one prepares a probe system in a suitably chosen state, lets it evolve according to the critical Hamiltonian, and finally measures it. This bear close similarity to the standard interferometric paradigm of quantum metrology \cite{demkowicz-dobrzanski_chapter_2015}.
On the other hand,  the "static" approach~\cite{zanardi_quantum_2008,bina_dicke_2016} is based on the equilibrium properties of the system. It consists in preparing and measuring the system ground state in the unitary case, or the system steady-state when open quantum systems are considered. In proximity of the phase transition the susceptibility of the equilibrium state diverges, and so it does the achievable measurement precision. Unfortunately, the time required to prepare the equilibrium state diverges as well, both in the unitary~\cite{sachdev2007} and in the driven-dissipative case~\cite{Macieszczak2016,Minganti2018}, a behavior called critical slowing down. Only very recently, it has been demonstrated that for a large class of spin models these two approaches are formally equivalent~\cite{rams_at_2018}, and that they both make it possible to achieve the optimal scaling limit of precision  with respect to system size and to measurement time. These results were obtained considering spin systems that undergo quantum phase transitions in the thermodynamic limit, where the number of constituents goes to infinity.
Another interesting class of quantum critical systems is provided by light-matter interaction models~\cite{Kirton2019}, for which superradiant quantum phase transitions can be controllably implemented~\cite{Zhiqiang2017,Baumann2010}. Recently, it has been  theoretically shown that quantum phase transitions can appear also in quantum-optical systems with only a finite number of components, where the thermodynamic limit can be replaced by a scaling of the system parameters\cite{bakemeier2012quantum,Ashhab2013, hwang_quantum_2015, peng_unified_2019, felicetti2019universal}. 

In this letter, we assess the metrological potential of a quantum phase transition taking place in a finite-component quantum optical model. More specifically, we design parameter-estimation protocols based on equilibrium properties of the quantum Rabi model, which exhibits a superradiant phase transition despite involving only one spin interacting with a bosonic field. 
In order to make a fair comparison with relevant benchmark  protocols, we explicitly take into account the time needed to prepare the ground state and the steady state, in the unitary and driven-dissipative case, respectively. We find analytical expressions for the scaling of the quantum Fisher information, and we find that this approach allows one to measure both spin and bosonic frequency with a favourable time scaling, in spite of the critical slowing down. In particular, we show that for spin frequency estimation our protocol exhibits time-scaling advantage with respect to the paradigmatic Ramsey protocol, while for bosonic frequency estimation it saturates the Heisenberg limit.

\paragraph{Protocol} 
Let us consider a spin interacting with a single bosonic mode according to the quantum Rabi Hamiltonian:
\begin{equation}
\label{RabiH}
	\hat{H} = \omega_0 \  \adag \aop + \Omega \Sz + \lambda\left(\adag + \aop \right) \Sx
\end{equation}
where $\omega_0$ is the frequency of the bosonic field, $\aop$ and $\adag$ are creation and annihilation operators of the field, $\Sx$ and $\Sz$ are Pauli matrices associated with the spin, and $\lambda$ is the coupling parameter. We also define the renormalized coupling parameter $g=\lambda/\sqrt{\Omega\omega_0}$. In the limit $\eta=\omega_0/\Omega\rightarrow0$, this system exhibits a phase transition at $g=1$ \cite{hwang_quantum_2015,peng_unified_2019}. 
We will analyze different critical quantum-metrology protocols that make use of this phase transition to estimate either the spin ($\Omega$) or the field ($\omega_0$) frequency, assuming in each case that all other parameters are known. 
In particular, we consider the following three-steps protocol: first, the system is initialized in its ground state for $g=0$; then, an adiabatic sweep is performed varying the parameter $g$ from 0 to some desired value close to the critical point $g=1$; finally, the measurement of a relevant observable is performed. The measurement results can then be used to estimate the desired parameter. 
In order to evaluate the performances of these protocols, we need first to characterize the system ground state as a function of the system parameters. In the limit $\eta\rightarrow0$, the system can be diagonalized using a Schrieffer-Wolff transformation \cite{hwang_quantum_2015}. We apply the unitary $\hat{U}=e^{ig\sqrt{\eta}(\adag+\aop)\Sy}$ to \eqref{RabiH}, which gives ${\hat{H}}^N=\hat{U}\hat{H}\hat{U}^{\dagger}$, where
\begin{equation}
\label{Hnormalphase}
{\hat{H}}^N=\omega_0 \adag\aop + \Omega\Sz + \frac{\omega_0}{2} g^2 
\Sz \Round{\aop + \adag}^2 \,,
\end{equation}
up to terms $O(\omega_0\sqrt{\eta})$. The effective Hamiltonian ${\hat{H}}^N$ provides a faithful description of the system ground state in the normal 
phase of the model. It is stable for $g<1$, whereas for $g\rightarrow 1$ 
the system experiences a phase transition towards the superradiant phase. Here, we will focus on the normal phase only, however equivalent results can be found applying the same methods to the superradiant phase (See supplementary material). In the normal phase, we can diagonalize ${H}^N$ by projection in the lower spin eigenspace 
and Bogoliubov transformation. The ground state is given by
\begin{equation}
\label{Groundstateadiab}
	\ket{\Psi_N}(\lambda,\Omega,\omega_0)=\hat{S}(\xi)\ket{0}\otimes\ket{\downarrow} 
\end{equation}
up to terms $O \left(\sqrt{\eta}\right)$. In Eq. (\ref{Groundstateadiab})
$\xi=-\frac{1}{4}\log(1-g^2)$ and ${\hat{S}}(\xi)=\exp\{\frac{\xi}{2}(\adag)^2 -\frac{\xi^*}{2} \aop\}$ is a squeezing operator. The squeezing parameter
diverges at the critical point, whereas the spin fluctuations are negligible, due to the much larger spin frequency. In turn, the excitation energy $\epsilon_N=\omega_0\sqrt{1-g^2}$ vanishes at the transition.

\begin{figure}
\includegraphics[width=0.5\columnwidth]{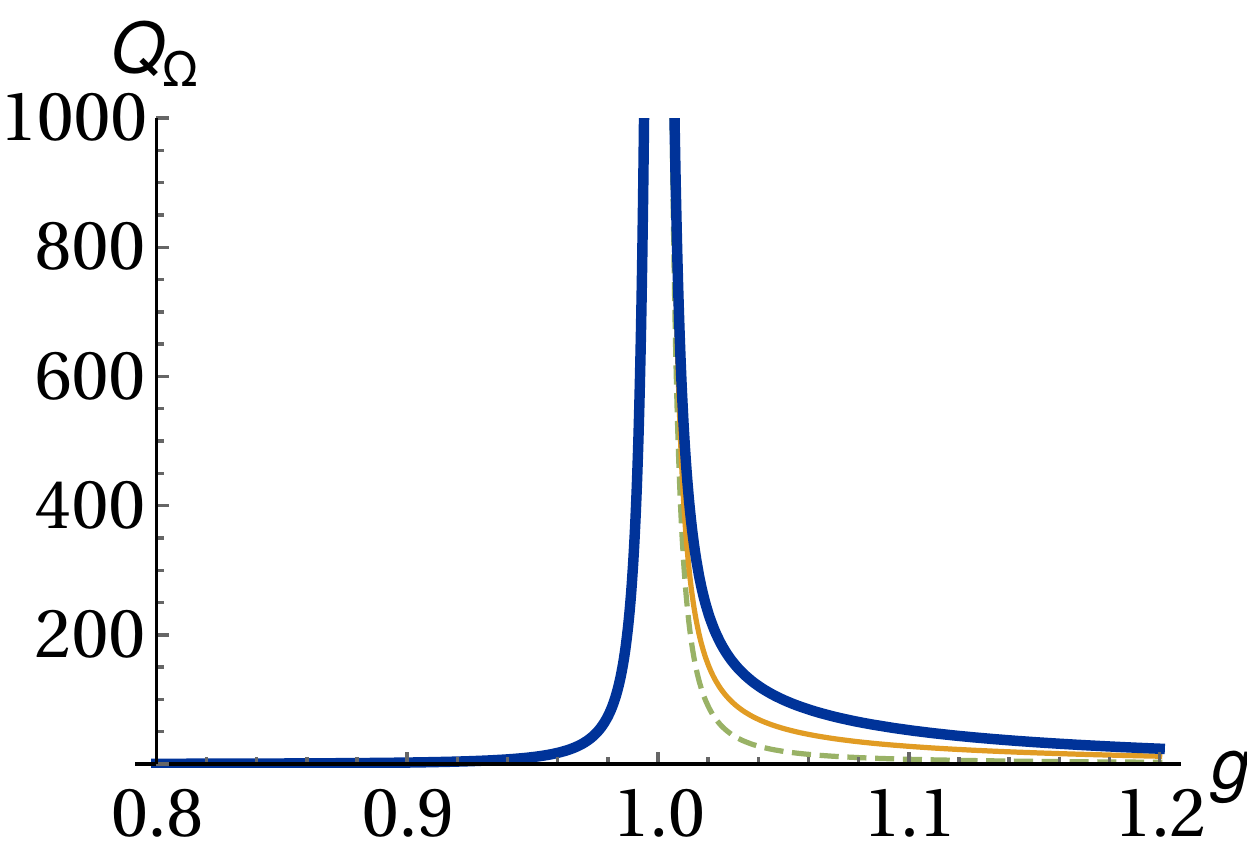}
\includegraphics[width=0.49\columnwidth]{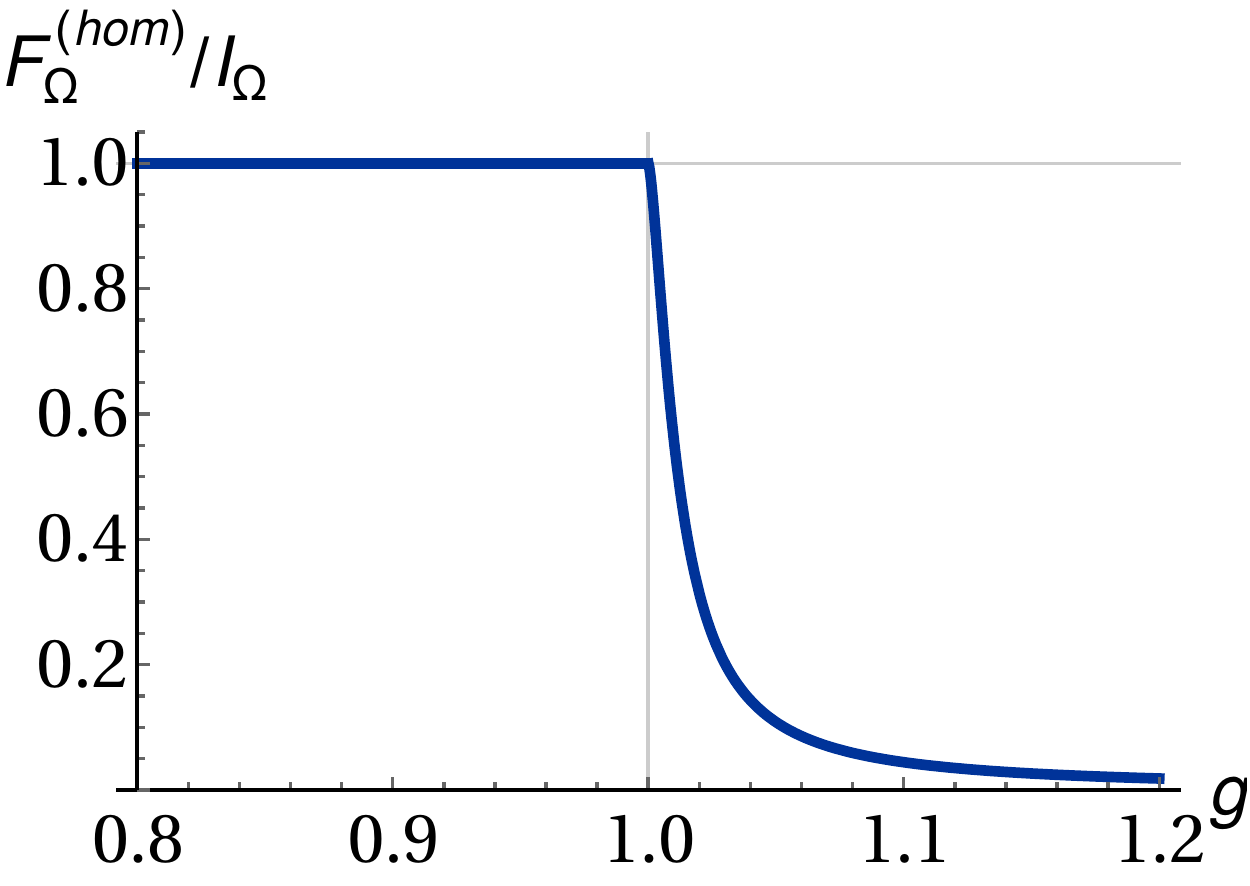}
\label{Fig}
\caption{Left: signal-to-noise ratio versus $g$, for $\Omega/\omega_0=10$ (thin dashed line), $50$ (thin full line), and $100$ (thick line). In the normal phase, the signal-to-noise ratio is independent of $\omega_0/\Omega$ for the value considered. In the superradiant phase, there is a small correction which becomes negligible near the critical point. Right: ratio FI/QFI for homodyne measurement of the $\frac{\xop+\pop}{\sqrt{2}}$ quadrature. In the normal phase $g<1$, the Cramer-Rao bound is attained for all values of $g$.
\label{Fig1}}
\end{figure}

We are interested in the precise estimation of $A$ (with $A=\Omega$ or $\omega_0$) obtained by performing measurements on the ground state of the
system. This precision is bounded by the quantum Cramer-Rao (CR) 
bound: $\delta^2 A\geq \mathcal{I}_{A}^{-1}$, where $\mathcal{I}_{A}$ is the Quantum Fisher Information (QFI) relative to the parameter of interest $A$. Since the system is in a pure 
state, the QFI may be computed exactly as $\mathcal{I}_{A}=4[\langle\partial_{A}\psi_{N}|\partial_{A}\psi_{N}\rangle+(\langle\partial_{A}\psi_{N}|\psi_{N}\rangle)^2]$. The dominant term of the QFI is: 
\begin{equation}
\label{QFIadiab}
	\mathcal{I}_A \simeq \frac{1}{32\,A^2(1-g)^2}\,,
\end{equation}
which means that the estimation of $\omega_0$ and $\Omega$ will yield the same signal-to-noise ratio $Q_A = A^2 \mathcal{I}_A$. Eq. (\ref{QFIadiab}) shows that $\mathcal{I}_A$
diverges at the critical point $g=1$, i.e. an arbitrarily-large estimation precision could in principle be obtained.
This is  consistent with previous studies on critical metrology in light-matter systems \cite{bina_dicke_2016}. To verify whether this bound is saturable with practical observables, we have also studied the Fisher information (FI) of
a feasible measurement, i.e. homodyne detection on the field only. 
This is illustrated in Fig. \ref{Fig1}, where we show $Q_\Omega$ 
versus $g$ for different values of the ratio $\Omega/\omega_0$ (left panel), and the ratio FI/QFI for homodyne detection of the $\frac{\xop+\pop}{\sqrt{2}}$ quadrature (right panel). In the normal phase, homodyne measurement allows to saturate the Cramer-Rao bound for all values of $g$. We found that other quadratures, such as $x$, also allows to saturate the Cramer-Rao bound. 

\paragraph{Analysis of resources} 
Let us now assess the performances of the proposed method taking standard metrological protocols as a benchmark. 
For the estimation of the bosonic frequency $\omega_0$, 
this is provided by interferometric protocols involving a phase 
difference $\Delta\phi=\omega_0\,T$ where $T$ is the evolution 
time within the interferometer. To ensure 
a fair comparison, we must carefully account for the resources needed to implement the critical and interferometric protocols. The relevant quantities to be considered are the evolution time $T$ and the average number of photons 
involved $\langle {}{N} \rangle$. A lossless interferometric protocol 
have a precision limited by the Heisenberg limit
$\mathcal{I}_{\omega_0}\sim \langle N \rangle^2\,T^2$. 
For the proposed critical protocol, we can readily compute $\langle N \rangle$
using Eq. \eqref{Groundstateadiab} as
$\langle \psi_N \lvert 
N \rvert \psi_N\rangle=\sinh{\xi}^2\simeq\frac14 (1-g^2)^{-\frac12}$ . 
Regarding the duration $T$ of the protocol, the relevant contribution is given by the time required to perform the adiabatic evolution. Since the gap closes at the critical point, the adiabatic evolution speed needs to be reduced in order to get closer to this point. This means that the time needed to reach a point arbitrarily close 
to the transition diverges. To estimate this time, we considered a general adaptative process during which $g$ evolves with a speed $v(g)=dg/dt$. We do not require the process to be a linear ramp, therefore $v$ can depend on $g$ in an arbitrary way. Using adiabatic evolution theory (see the Supplemental Material), we look for optimal adiabatic procedures that minimize the evolution time while ensuring that the system will remain in the ground state during the evolution. We find the following condition on the speed $v$  of evolution,
\begin{equation}
	v(g)\sim\gamma\omega_0\,(1-g^2)^{3/2}\,.
\end{equation}
where $\gamma<1$ is a parameter which controls the probability of exciting the system.
As a result, the time needed to sweep the coupling constant from 
$0$ to some value $g\simeq1$ is given by
 \begin{equation}
 \label{adiabtime}
 	T=\int_0^g\!\! \frac{ds}{v(s)} \sim \gamma^{-1}\omega_0^{-1}\, (1-g)^{-\frac12}\,,
 \end{equation}
This expression indeed diverges when $g$ goes to $1$. Upon inserting the 
expressions for $\langle {}{N} \rangle$ and $T$ into \eqref{QFIadiab}, 
we find
\begin{equation}
\label{scalingomega0adiab}
	\mathcal{I}_{\omega_0}\sim \gamma^2 \langle N \rangle^2\, T^2\,,
\end{equation}
i.e., the critical protocol allows one to estimate $\omega_0$ with 
the same precision granted by interferometric protocols. In other words, in spite of the critical slowing down the critical protocol achieves the optimal Heisenberg-scaling precision for continuous-variable systems, with respect to both energy and time. Similar results had been obtained for critical spin systems in the thermodynamic limit\cite{rams_at_2018}.
Concerning the estimation of the spin frequency, a natural benchmark is given by Ramsey interferometry with a single spin. For noiseless Ramsey 
interferometry, QFI scales like $T^2$ \cite{huelga_improvement_1997,giovannetti_quantum-enhanced_2004}. By contrast, in the critical 
case, we found using $\eqref{QFIadiab}$ and $\eqref{adiabtime}$:
 \begin{equation}
 \label{scalingadiab}
 	\mathcal{I}_{\Omega}\simeq\frac{\gamma^4\omega_0^4}{8\Omega^2}T^4\,,
 \end{equation}
i.e. our protocol achieves quartic scaling in the duration of the protocol, while Ramsey interferometry only scales quadratically. To the best of our knowledge, this is the first unambiguous demonstration of time-scaling 
advantage for a  critical metrological protocol in light-matter system. Note however that the prefactor in \eqref{scalingadiab} is very small, meaning that the critical protocol could outperform Ramsey only for large measurement time $T$.

\paragraph{Dissipative process} The above results are valid for isolated 
systems. However, decoherence due the interaction with the environment, generally reduces the performances of metrological protocols.  In order 
to assess our protocol in realistic conditions, let us now consider 
the presence of both photon loss and spin decay. The dissipative dynamics 
of the system is described by a master equation (ME) of the form
\begin{equation}
\label{lindblad}
\dot{\hat{\rho}}=-i[\hat{H},\hat{\rho}]+\kappa L[\aop]\rho+
\Gamma L[\Sm]\rho\,,
\end{equation}
where the Lindblad terms read $L[\hat{A}]\rho=2\hat{A}\hat{\rho}\hat{A}^{\dagger}-({\hat{A}^{\dagger}\hat{A},\hat{\rho}})$.  Notice that we are considering a phenomenological master equation as we are interested in effective implementations of the model~\cite{Puebla2017}.  To characterize the dissipative case we will generalize the results obtained in \cite{hwang_dissipative_2018} to include spin decay, details can be found in the Supplemental Material. We then assume $\kappa/\omega_0=O(1)$ and  $\Gamma/\Omega=O(1)$, however our results can be readily extended to a broader regime of parameters (for instance when $\Gamma=O(\sqrt{\omega_0\Omega})$). Upon considering 
the spin-decay term explicitely and using Schrieffer-Wolff transformation, 
we decouple the spin and field, and project the spin into the $\ket{\downarrow}\bra{\downarrow}$ subspace. This yields an effective ME for the bosonic part
\begin{eqnarray}
\dot{\hat{\rho}}_b=-i[\omega_0 \adag\aop - Y   (\aop + \adag)^2,\hat{\rho}_b]+ \nonumber \\
+\kappa L[\aop](\hat{\rho}_b)+ \frac{\Gamma}{\Omega}\,Y L[\aop+\adag]\hat{\rho}_b,
\end{eqnarray} 
plus terms of order $O(\omega_0\sqrt{\eta})$. We defined $X=\Omega^2/(\Gamma^2+\Omega^2)$ and $Y=\frac14 \omega_0 X{g}^2$. 
Since this equation is quadratic in $\aop$, it can be solved by a Gaussian ansatz. The dynamics is then fully characterized by the evolution 
equation for the covariance matrix $\sigma$ of the state. The displacement 
vector decays quickly to zero and may be safely discarded, so we obtain
$\partial_t\sigma   = B\sigma+\sigma B^T- 2\kappa(\sigma - \sigma_{\hbox{\tiny L}})$ where
\begin{align}
\label{equationsigma}\nonumber
B  & =\begin{pmatrix}0 & \omega_0\\ 4 Y - \omega_0 & 0\end{pmatrix},
\end{align}
and
$\sigma_{\hbox{\tiny L}} = \frac12 [{\mathbb{I}}+\hbox{Diag}(0,4Y \Gamma/(\Omega\kappa))]$.
This linear equation may be solved exactly by diagonalization. Upon 
evaluating the lowest eigenvalue, one may estimate the typical time 
needed to reach the steady-state, $T\simeq g_c/\kappa\, (g-g_c)^{-1}(1+\omega_0^2/\kappa^2)^{-1}$. This value diverges near the transition, 
indicating a critical slowing down. The steady-state is a squeezeed 
(undisplaced) thermal state, with covariance matrix given by
\begin{equation}
\label{sigma}
	\sigma=\frac{1}{2}\, \mathbb{I} + \frac{g^2\left(1+\frac{\omega_0\Gamma}{\Omega\kappa}\right)}{4(g_c^2-{g}^2)}\begin{pmatrix} 1 & \frac{\kappa}{\omega_0} \\ \frac{\kappa}{\omega_0} & X{g}^2-1\end{pmatrix}
\end{equation} 
with $g_c^2=(1+\Gamma^2/\Omega^2)(1+\kappa^2/\omega^2)$. In this dissipative setting, the system still experiences a phase transition for $g\rightarrow g_c$. Both the squeezing and thermal energies of the steady-state diverge near the critical point. 
Since this state is Gaussian and its first-moment vector is zero, 
the QFI may be evaluated as (dots denote derivative with respect to the
parameter under consideration)
$$
\mathcal{I}_A=\frac{8}{16\,d^4-1}\left\{d^4\,\text{Tr}\left [ (\sigma^{-1}\dot{\sigma})^2 \right ]-\frac14\, \text{Tr}\left [ (\dot{\sigma}\,\omega)^2 \right ]\right\}\,,$$
with $d=\sqrt{\text{Det}\,\sigma}$ \cite{monras_phase_2013}. The leading
terms of the QFIs for the estimation of frequencies are given by 
\begin{align}\label{scalingdiss}
\mathcal{I}_{\omega_0}^{\text{diss}}\simeq\frac{2\Omega}{\Omega\kappa+\omega_0\Gamma}\left(\frac{\kappa^2-\omega_0^2}{\kappa^2+\omega_0^2}\right)^2\langle N \rangle T, \\ \nonumber
 \mathcal{I}_\Omega^{\text{diss}}\simeq\left(\frac{\Gamma^2-\Omega^2}{\Gamma^2+\Omega^2}\right)^2\frac{\kappa^2}{\Omega^2}\left(1+\frac{\omega_0^2}{\kappa^2} \right)^2T^2\,.
\end{align}
Eq. (\ref{scalingdiss}) shows that for the estimation of $\omega_0$, the 
presence of dissipation restores the shot-noise scaling, similar to what happens in a lossy interferometric protocols. 
In the case in which the parameter to be estimated is the spin frequency $\Omega$, the presence of dissipation replaces the quartic time-scaling obtained in the Hamiltonian case \eqref{scalingadiab} by a quadratic one. However, the QFI of a Ramsey protocol in presence of spin decay at rate 
$\Gamma$ is given by $\mathcal{I}_\Omega=T/\Gamma$, and so it is linear in time. This result shows that
the time-scaling advantage of our critical protocol against the benchmark   persists in the dissipative case for spin-frequency estimation.

\paragraph{Discussion} 
Let us now comment on the nature, the limitations and the potential experimental implementations of the considered protocols. First of all, we emphasize that our protocol exploits the diverging susceptibility near the transition, but it does not require to actually cross the critical point, contrary to what is used in transition-edge sensors \cite{irwin_transition-edge_2005}.
Besides, in contrast to the standard interferometric setting of quantum metrology, in our scheme the preparation and the phase acquisition stages are performed together. Overall, our metrological protocol corresponds to a squeezing channel applied to an initial vacuum state. Accordingly, the estimation of the bosonic frequency $\omega_0$ amounts to evaluating the  squeezing parameter of this channel, and indeed it achieves the optimal Heisenberg scaling. Concerning the estimation of the spin frequency $\Omega$, our critical protocol achieves time-scaling advantage compared to Ramsey protocol. However, since the prefactor in \eqref{scalingadiab} and \eqref{scalingdiss} is small, the critical protocol outperforms Ramsey schemes only for long protocol duration, i.e.  when operating in close proximity of the critical point. In this region, the quartic and higher-order terms in the Schrieffer-Wolff expansion of the Hamiltonian, that we have neglected in order to obtain exact results, may become relevant~\cite{hwang_quantum_2015}. As a consequence, the exact point at which our critical protocol will outperform standard Ramsey protocol is difficult to evaluate and will depends on the details of the  experimental implementation. 

Concerning possible experimental realizations, analog quantum simulation techniques have been applied to implement the quantum Rabi model in extreme regimes of parameters using different quantum technologies, such as cold atoms \cite{Dareau2018}, superconducting circuits \cite{braumuller2017} and trapped ions \cite{Lv2018}. Finite-component driven-dissipative phase transitions can  be implemented with bath-engineering techniques~\cite{Puebla2017}. Furthermore, it has been recently shown~\cite{felicetti2019universal} that finite-component phase transitions can be observed also with weakly-anharmonic quantum resonators, so our results could be extended to include nonlinear quantum resonator  implemented with circuit-QED devices \cite{Markovic2018} and electromechanical systems \cite{Peterson2019}.

\paragraph{Conclusions and outlook} 
Our results show that, in spite of the critical slowing down, critical quantum-optical systems represent a compelling tool for quantum metrology. Furthermore, we have demonstrated the metrological potential of finite-component quantum phase transitions, a result that have both practical and fundamental consequences. Finite-component criticalities allow us to substantially reduce the system size and complexity, at the cost of accessing an unusual regime of parameters. A promising perspective consists in the application  of quantum-control schemes to reduce the time required to perform an adiabatic sweep in critical quantum metrology. Indeed, in a finite-component system quantum-control techniques could be applied without implementing complex non-local operations, as it is the case for many-body systems. 
In addition, our study paves the way to the application of other criticalities appearing in quantum-optical models~\cite{Bartolo2016Exact,Casteels2017Critical,garbe_superradiant_2017, felicetti2019universal} in quantum metrology. Finally, by focusing on the time-scaling and on a finite-component system, our analysis challenges the standard framework in which the fundamental resources needed to achieve metrological quantum advantage are assessed~\cite{Hyllus2010,Yadin2018,Kwon2019,Garbe2019}.


\paragraph{} -- We thank Philipp Schneeweiss for useful discussions. S. F. acknowledges support from the European Research Council (ERC-2016-STG-714870).

\bibliographystyle{apsrev4-1}
\bibliography{References.bib}

\newpage

\begin{widetext}

\begin{center}\textbf{Supplementary material}\end{center}

This Supplementary material is composed of several sections: first, we compute the ground state of the Rabi model in the superradiant phase and discuss its use for quantum metrology. Next, we present the technical details related to the adiabatic process. Finally, the last section provides the formal treatment of the dissipative case.

\section{Metrology in the superradiant phase}
In the superradiant phase, the field quadratures acquire a non-zero mean value. To take this into account, we apply a displacement operator $\hat{D}(\alpha)=\exp\{\alpha \adag -\alpha^* \aop \}$ (with $\alpha$ real) to the Hamiltonian. Two possible values of $\alpha$, $\alpha=\pm\alpha_s=\pm\frac{1}{2g}\sqrt{\frac{\Omega}{\omega_0}}\sqrt{g^4-1}$, will give stable dynamics:

\begin{equation}
\hat{H}(\pm\alpha_s) = \hat{D}^\dagger(\pm\alpha_s) \hat{H}  \hat{D}(\pm\alpha_s) = \omega_0 \adag \aop \pm \omega_0 \alpha_s \Round{\adag + \aop} + \omega_0g\sqrt{\frac{\Omega}{\omega_0}} \Round{\adag + \aop} \Sx + \Omega \Sz  \pm 2\alpha_s\omega_0g\sqrt{\frac{\Omega}{\omega_0}} \Sx + \text{const}.
\end{equation}

We can rewrite this Hamiltonian in a new spin basis $\hat{\tau}_i^{\pm}=e^{\Curly{\mp \imm \theta \Sy }} \hat{\sigma}_i e^{\Curly{ \pm \imm \theta \Sy }}$, where $\tan\Round{\theta} = 2g\alpha_s\sqrt{\frac{\omega_0}{\Omega}}$. Then we perform again Schrieffer-Wolff treatment, by applying $$\hat{U}=\exp\Curly{\frac{\imm}{g^3}\sqrt{\frac{\omega_0}{\Omega}}(\aop+\adag)\Ty^{\pm}-\frac{\imm}{g^6}\frac{\omega_0}{\Omega}\sqrt{g^4-1}(\aop+\adag)^2\Ty^{\pm}}$$ and projecting within the lower eigenspace of $\Tz^{\pm}$. This yields a field Hamiltonian $$\hat{H}(\pm\alpha_s)=\omega_0\adag\aop + \Omega g^2\Tz^{\pm} \pm\omega_0\alpha_s(\adag+\aop)[1+\Tz^{\pm}] + \frac{\omega_0}{2g^4}(\adag+\aop)^2\Tz^{\pm}$$ 
 The ground-state of this Hamiltonian is a squeezed state with squeezing parameter $\xi^S=-\frac{1}{4}\log(1-\frac{1}{g^4})$, and the excitation energy is $\epsilon_S=\omega_0\sqrt{1-\frac{1}{g^4}}$. Thus, we arrive at the two following states: $ \ket{\Psi_S^{\pm}}=\hat{D}(\pm\alpha_s){\hat{S}}(\xi^{S})\ket{0}\otimes\ket{\downarrow^{\pm}}$ with $\ket{\downarrow^{\pm}}$ the lower eigenstate of $\tau_z^{\pm}$. In the dissipative case, we will obtain two degenerate steady-state; this is a generic property of symmetry-breaking phase transitions \cite{Minganti2018}.

 For both $\ket{\Psi_S^{+}}$ and $\ket{\Psi_S^{-}}$, we can compute directly the QFI and the time needed for adiabatic evolution. We find $\mathcal{I}_{A}\sim \frac{1}{2A^2}\frac{1}{(g^4-1)^2} + \frac{1}{\Omega\omega_0}\frac{1}{g^4\sqrt{g^4-1}}$ for $A=\omega_0,\Omega$. Notice that the second term becomes negligible when $g$ goes to $1$. Finally, this yields once more: $\mathcal{I}_{\omega_0}\sim\langle N \rangle^2T^2$ and $\mathcal{I}_{\Omega}\sim\frac{\omega_0^4}{\Omega^2}T^4$.
  In the dissipative case, however, we expect that the preparation procedure will create a mixture of the two symmetry sector. Although it becomes challenging to compute the QFI exactly in that case, we expect this mixture to reduce the performances of the protocol. 

\section{Requirements for adiabatic process.}

In the normal phase, we let the coupling constant $g$ evolve (in general non-linearly) in time. The dynamics of the system is described by the time-dependent Hamiltonian $\hat{H}^N(t)=\omega_0\adag\aop +\Omega\hat{\sigma}_z+\frac{\omega_0}{2}g(t)^2\hat{\sigma}_z(\aop+\adag)^2$. The instantaneous eigenstates $\ket{n_s(t)}$ are given by squeezed Fock states,
\begin{equation}
\hat{H}(t) \ket{n_s(t)}= n \epsilon(t) \ket{n_s(t)},\quad \end{equation}
\begin{equation} \ket{n_s(t)} = S(t) \ket{n} = \exp\Curly{\frac{\xi(t)}{2}\Round{({\adag})^2 - {\aop}^2  }}\ket{n},
\end{equation}
with $\xi(t) = - \frac{1}{4}\log\Square{1-g^2(t)} $, and the energy gap is given by $\epsilon(t) = \omega_0 \sqrt{1-g^2(t)}$.
The system state can then be decomposed over this basis as:
\begin{equation}
\ket{\psi(t)} = \sum_n \alpha_n(t) e^{-i\Theta_n(t)} \ket{n_s(t)}\eeq where
\beq \Theta_n(t)= \int_0^t n \epsilon(t^\prime) dt^\prime,
\end{equation}
The goal is then to maintain the system in the ground state, ie, $\alpha_n=0$ for $n\neq0$. We can compute the evolution of the $\alpha_n$ coefficients by using the Schrödinger equation:
\begin{equation}
\frac{d\alpha_n(t)}{dt} = - \sum_m \alpha_m(t) e^{-i\Square{\Theta_m(t) - \Theta_n(t) } } \bra{n_s(t)} \frac{\partial}{\partial t}\ket{m_s(t)},
\end{equation}
which we can formally solve in time, and which we rewrite changing the integration variable using $\delta g = v \delta t$
\begin{equation}
\alpha_n(g) = 
 -\sum_m \int_0^g \alpha_m(g^\prime) e^{-i\Square{\Theta_m(g^\prime) - \Theta_n(g^\prime) }  }  \bra{n_s(g^\prime)}\frac{\partial}{\partial g^\prime}\ket{m_s(g^\prime)}.
\end{equation}

We assume that the system is initially in its ground state, $\alpha_m(0) = 1$ for $m=1$ and $\alpha_m(0) = 0$ otherwise. Time-dependent perturbation theory allows us to write:
\begin{equation}
\alpha_n(g) = -\int_0^g e^{-i\Square{\Theta_0(g^\prime) - \Theta_n(g^\prime) }  }  \bra{n_s(g^\prime)}\frac{\partial}{\partial g^\prime}\ket{0_s(g^\prime)} 
\end{equation}
We can calculate directly the matrix element $\bra{n_s(g^\prime)}\frac{\partial}{\partial g^\prime}\ket{0_s(g^\prime)}$,
\begin{equation}
\label{overlapnondiss}
\bra{n_s(g^\prime)}\frac{\partial}{\partial g^\prime}\ket{0_s(g^\prime)}=\bra{n}\hat{S}^\dagger(g^\prime)\frac{\partial}{\partial g^\prime}\hat{S}(g^\prime)\ket{0} = \frac{\sqrt{2}}{4}\frac{g^\prime}{1-{g^\prime}^2} \delta_{n,2},
\end{equation}

so at the order considered only transitions to the second-excited state $S(g^\prime)\ket{2}$ should be taken into account. Thus we can rewrite,
\begin{eqnarray}
\label{eqalpha}
\alpha_2(t) & = -\frac{1}{2\sqrt{2}}\int_0^g  \frac{g^\prime}{1-{g^\prime}^2} e^{iR(g^\prime)} dg^\prime,\\ \nonumber
 & = -\frac{1}{2\sqrt{2}}\int_0^gf(g^\prime)e^{iR(g^\prime)}dg^\prime
\end{eqnarray}
where we defined $f(g)=\frac{g}{1-{g}^2}$ and
$R(g) = \Theta_2(g) - \Theta_0(g) = 2\omega_0\int_0^g  \frac{\sqrt{1- {g^\prime}^2}}{v(g^\prime)} dg^\prime.$

We now want to choose $v$ to ensure that $\alpha_2$ remains small during the evolution. We will first propose an ansatz for the speed based on an hand-waving argument, then we will justify formally that this expression gives the desired results.

To have an adiabatic sweep, we need $v(g)\ll1$ so that $R(g)$ is large and the exponential in the integral in Eq.~\eqref{eqalpha} oscillates fast, cancelling the integral. More precisely, the exponential term must oscillate faster than the evolution of $f$; otherwise the changes of $f$ have the possibility to build up before the oscillations can kill them. This suggest that our goal will be reached when the following condition is satisfied: $\frac{\dot{f}}{f}\ll\dot{R}$, where here the dot means derivative with respect to $g$. We expect that the behavior of the system will be dominated by the end of the evolution, when $g$ gets close to $1$. Therefore we may set $f(g)\sim(1-g^2)^{-1}$.  Hence, we  arrive at the following ansatz for the speed:

\begin{equation}
\label{ansatzv}
	v(g^{\prime})=\gamma\omega_0(1-{g^{\prime}}^2)^{3/2}
\end{equation}
with $\gamma$ a small constant. We will now show that this expression gives the desired results. We rewrite Eq.~\eqref{eqalpha} using integration by part:
\begin{align}
\nonumber
\alpha_2(t) &= +\frac{i}{2\sqrt{2}}\int_0^g  \frac{g^\prime}{\Round{1-{g^\prime}^2}\dot R(g^\prime)} \frac{\partial}{\partial g^\prime} e^{iR(g^\prime)} dg^\prime \\  & =\frac{i}{4\sqrt{2}\omega_0}\Big(\frac{v(g) g  e^{iR(g)} }{\Round{1-g^2}^{3/2}} -
\int_0^g  e^{iR(g^\prime)} \frac{\partial}{\partial g^\prime} \Square{\frac{v(g^\prime)g^\prime}{\Round{1-{g^\prime}^2}^{3/2}}} dg^\prime\Big)
\end{align}

Using \eqref{ansatzv}, the second term can be rewritten as (up to unimportant constant factors):
 \begin{align}\nonumber
 	 \alpha_2^{(1)} & =\int_0^g \gamma e^{iR(g^\prime)}\ll\int_0^gf(g^\prime)e^{iR(g)}
 \end{align}
 since $\gamma\ll1$. Hence the second term is negligible with respect to the complete factor $\alpha_2$ and can be omitted.
 Finally, we can write the probability of exciting the system during the adiabatic sweep as
\begin{equation}
\left| \alpha_2 (g) \right|^2 = \frac{v(g)^2}{32\omega_0^2}\frac{g^2}{\Round{1-g^2}^{3}}. 
\end{equation}
Which is indeed small with respect to $1$. We have thus proven the validity of the ansatz \eqref{ansatzv}, and we can now use it to evaluate the total time needed to perform the adiabatic evolution:

$$T=\int_0^{g}\frac{1}{v(g^{\prime})}dg^{\prime}\rightarrow\frac{1}{\gamma\omega_0(1-g)^{1/2}}\sim\frac{1}{\gamma\epsilon_N(g)}$$

\par
\section{Dissipative dynamics.}
We will now describe the dynamics of the system under the Lindblad equation \eqref{lindblad} given in the main text. For convenience, we will assume $\Gamma=O(\Omega)$ and $\kappa=O(\omega_0)=O(\Gamma\eta)$, however our results can be extended to the case $\Gamma=O(\sqrt{\Omega\omega_0})$ with only minor adjustements. We apply the transformation $\hat{U}=e^{ig\sqrt{\eta}(\adag+\aop)\Sy}$, which yields:

\begin{align}
\label{lindblad_complet}
  	\dot{\rop}= & -i[\hat{H}^N,\rop]+\kappa L[\aop](\rop)+\Gamma L[\Sm](\rop) \\ \nonumber
  	& + 2\Gamma g \sqrt{\eta}\left((\aop+\adag)\Sz\rho\Sp +h.c.\right) +\Gamma g \sqrt{\eta}\Curly{(\aop+\adag)\Sx,\rop}\\ \nonumber
  	& -g^2\Gamma\eta\left((\aop+\adag)^2\Sx\rop\Sp+h.c.-2(\aop+\adag)\Sz\rop\Sz(\aop+\adag)\right) + \frac{\Gamma g^2 \eta}{2}\Curly{(\aop+\adag)^2\Sz,\rop}+O(\omega_0\sqrt{\eta}) 
  	  \end{align}
where $H_N$ is the Hamiltonian in the normal phase \eqref{Hnormalphase}, and where we have defined $\eta=\frac{\omega_0}{\Omega}$. First, we will focus on the spin dynamics. We decompose the state $\rop$ into its spin components:
\begin{equation}
	\rop=\rop_{bd}\ket{\downarrow}\bra{\downarrow} + \rop_{bu}\ket{\uparrow}\bra{\uparrow} + \rop_{bc}\ket{\downarrow}\bra{\uparrow}+ \rop^{\dag}_{bc}\ket{\uparrow}\bra{\downarrow}
\end{equation}
where $\rop_{bd}$, $\rop_{bu}$, $\rop_{bc}$ and $\rop^{\dag}_{bc}$ are bosonic operators. In \eqref{lindblad_complet}, the term $\Gamma L[\Sm](\rop)$ will tend to bring us to the $\ket{\downarrow}\bra{\downarrow}$ subspace. The following terms, which create non-zero value outside this subspace, are only of order $\Gamma\sqrt{\eta}$ and $\Gamma\eta$. Therefore, we make the following ansatz for our four state components, which will be verified at the end of our analysis:
$$\rop_{bd}=O(1), \ \rop_{bc}=O(\sqrt{\eta}), \ \rop_{bu}=O(\eta)$$

Then by projecting \eqref{lindblad_complet}, we obtain:

\begin{align}
	\dot{\rop}_{bu}= & -2\Gamma\rop_{bu}+\frac{g\Gamma}{2}\sqrt{\eta}\left((\aop+\adag)\rop_{bc}+h.c.\right)+O(\Gamma\eta^2) \\
	\dot{\rop}_{bc}= & (i\Omega-\Gamma)\rop_{bc} + \frac{g\Gamma}{2}\sqrt{\eta}\rop_{bd}(\aop+\adag) +O(\Gamma\eta\sqrt{\eta}) \\ \nonumber
	\dot{\rop}_{bd}= & -i[\omega_0\adag\aop-\omega_0\frac{g^2}{4}(\aop+\adag)^2,\rop_{bd}] + 2\Gamma\rop_{bu} \\ \nonumber
	& + \kappa L[\aop](\rop_{bd}) -g\Gamma\sqrt{\eta}\left((\aop+\adag)\rop_{bc}+h.c.\right)\\ \nonumber
	& + \frac{g\Gamma}{2}\sqrt{\eta}\left((\aop+\adag)\rop^{\dag}_{bc}+h.c.\right) + \frac{g^2\Gamma}{2}\eta\left((\aop+\adag)\rop_{bd}(\aop+\adag)\right) \\
	& -\frac{g^2\Gamma}{4}\eta\Curly{(\aop+\adag)^2\rop_{bd}} + O(\Gamma\eta^2)
\end{align}
$\rop_{bu}$ and $\rop_{bc}$ evolve quickly, at a rate $\Gamma$. We will supress them using adiabatic elimination: this gives 
\begin{align}
	\rop_{bc}=\frac{g\Gamma}{2(\Gamma-i\Omega)}\sqrt{\eta} \rop_{bd} (\aop+\adag) + O(\eta\sqrt{\eta})\\
	\rop_{bu}=\frac{g^2}{4}\frac{\Gamma^2}{\Gamma^2+\Omega^2}\eta (\aop+\adag)\rop_{bd} (\aop+\adag) +O(\eta^2)
\end{align}

with these expressions, we have indeed $\rop_{bc}=O(\sqrt{\eta}), \ \rop_{bu}=O(\eta)$. Then we reinject these in the equation above, and obtain after straightforward calculations:
\begin{align}
	\nonumber \dot{\rop}_{bd}= & -i[\omega_0\adag\aop-\omega_0X\frac{g^2}{4}(\aop+\adag)^2,\rop_{bd}] + \kappa L[\aop](\rop_{bd})\\ 
	& + \frac{Xg^2}{4}\frac{\omega_0\Gamma}{\Omega} L[\aop+\adag](\rop_{bd})+O(\omega_0\eta)
\end{align}
with $X=\frac{\Omega^2}{\Gamma^2+\Omega^2}$. This equation describes the dynamics of the bosonic field inside the lower spin subspace.\\

 We will now move to phase space \cite{ferraro_gaussian_2005}
 and rewrite the Lindblad equation above into a Fokker-Planck equation for the Wigner function:
\begin{equation}
\frac{\partial W}{\partial t}(x,p)=-\omega_0p\frac{\partial W}{\partial x} - \omega_0(Xg^2-1)x\frac{\partial W}{\partial p}+\kappa(2W+x_i\partial_iW + \partial_i\sigma^L_{ij}\partial_jW),\end{equation}

Here $x_1=x$, $x_2=p$, we have used summation of repeated indices, and $$\sigma^L =\frac{1}{2}\begin{bmatrix}1 & 0 \\ 0 & 1+Xg^2\frac{\Gamma\omega_0}{\Omega\kappa}\end{bmatrix}$$
Since this equation is quadratic in $x$ and $p$, it can be solved by a Gaussian ansatz $W=\frac{1}{\sqrt{\pi \text{det}(\sigma)}}\exp\Curly{\frac{-1}{2}\textbf{x}^T\sigma^{-1}\textbf{x}}$. The displacement decays at a rate $2\kappa$ and will quickly reach $0$. Thus, this function is entirely caracterised by the covariance matrix, which is described by the following equation:

\begin{align}
\label{equationsigmabis}
\partial_t\sigma & = B\sigma+\sigma B^T- 2\kappa(\sigma - \sigma^L)
\end{align}
With:
\begin{align*}
B & =\begin{bmatrix}0 & \omega_0\\ \omega_0(Xg^2-1) & 0\end{bmatrix}
\end{align*}

The first term in \eqref{equationsigmabis} originates from the Hamiltonian dynamics. We will define the eigenmatrices of this evolution: $B M_i + M_i B^T=\lambda_iM_i$.
Before we give the expressions of $M_i$ and $\lambda_i$, let us emphasize that rigorously, we should distinguish between the cases $g^2\leq\frac{1}{X}$ and $g^2\geq\frac{1}{X}$. For $g\leq \frac{1}{X}$, the Hamiltonian is an ordinary squeezing Hamiltonian bounded from below, the $\lambda_i$ are complex, and the $M_i$ correspond to oscillating solution. For $g^2\geq\frac{1}{X}$, the Hamiltonian is no longer bounded from below, the $\lambda_i$ are real, and the $M_i$ are diverging (or vanishing) in time. Here we will focus only on the case $g^2\geq\frac{1}{X}$, however the formalism for the case $g^2\leq\frac{1}{X}$ is equivalent. We find the following eigenmatrices and eigenvalues:

\begin{center}
$M_0=\begin{bmatrix} 0 & 1 \\ -1 & 0\end{bmatrix}$, $M_1=\begin{bmatrix} \frac{1}{\sqrt{Xg^2-1}} & 0 \\ 0 & -\sqrt{Xg^2-1}\end{bmatrix}$, $M_{\pm}=\begin{bmatrix} \frac{1}{\sqrt{Xg^2-1}} & \pm1 \\ \pm1 & \sqrt{Xg^2-1}\end{bmatrix}$

\vspace{10pt}

$\lambda_0=\lambda_1=0$, $\lambda_{\pm}=\pm2\omega_0\sqrt{Xg^2-1}$.
\end{center}

The matrices $M_i$ form a complete basis, thus we may write $\sigma=\sum_i c_i M_i$ and $\sigma^L=\sum_i c^L_i M_i$: we find $c^L_0=0$, $c^L_1=\frac{Xg^2\left(1-\frac{\omega_0\Gamma}{\Omega\kappa}\right)-2}{4\sqrt{Xg^2-1}}$ and $c^L_\pm=\frac{Xg^2}{8\sqrt{Xg^2-1}}\left(1+\frac{\omega_0\Gamma}{\Omega\kappa}\right)$.
We can then rewrite \eqref{equationsigmabis} as an equation of evolution for the coefficients $c_i$:

\begin{align}
	\partial_tc_i & = (\lambda_i-2\kappa) c_i+2\kappa c_i^L\\
	& = -\tilde{\lambda}_i(c_i-\frac{2\kappa}{\tilde{\lambda}_i}c_i^L)
\end{align}
with $\tilde{\lambda}_i=2\kappa-\lambda_i$. This gives:

\begin{equation}
\label{citime}
c_i(t)=\Big(c_i(t=0)-\frac{2\kappa}{\tilde{\lambda}_i}c_i^L\Big)e^{-\tilde{\lambda}_it}+c_i^L\frac{2\kappa}{\tilde{\lambda}_i}
\end{equation}

We are finally able to find the steady-state of the system. First we compute the coefficients $c_i$ when $t\rightarrow \infty$. We find $c_i\rightarrow c_i^L\frac{2\kappa}{\tilde{\lambda}_i}$, and $\sigma=\sum_ic_i^L\frac{2\kappa}{\tilde{\lambda}_i}M_i$. Now, putting the expressions of $c_i^L$ and $M_i$, we find the covariance matrix \eqref{sigma}. 

Finally, we estimate the time needed to reach the steady-state. From \eqref{citime}, we see that the $c_i-c_i(t\rightarrow\infty)$ decay at various rates, the smallest one being $ \tilde{\lambda}_{+}=2\kappa-2\omega_0\sqrt{Xg^2-1}$, which near the transition is equal to $2\kappa\frac{g_c-g}{g_c}\left(1+\frac{\omega_0^2}{\kappa^2}\right)$. This vanishing decay rate will dominate the relaxation of the system near the critical point; hence, we can evaluate the duration of the protocol as $T\sim\frac{1}{2\kappa}\frac{g_c}{g_c-g}\frac{1}{1+\frac{\omega_0^2}{\kappa^2}}$.

\end{widetext}

\end{document}